# Superconductor-metal quantum transition at the EuO/KTaO$_3$ interface

Yang Ma[1†], Jiasen Niu[1†], Wenyu Xing[1], Yunyan Yao[1], Ranran Cai[1], Jirong Sun[2,3], X. C. Xie[1,4,5], Xi Lin[1,4,5]*, Wei Han[1]*

[1]International Center for Quantum Materials, School of Physics, Peking University, Beijing 100871, P. R. China

[2]Beijing National Laboratory for Condensed Matter Physics & Institute of Physics, Chinese Academy of Sciences, Beijing 100190, P. R. China

[3]School of Physical Sciences, University of Chinese Academy of Sciences, Beijing 100049, P. R. China

[4]CAS Center for Excellence in Topological Quantum Computation, University of Chinese Academy of Sciences, Beijing 100190, P. R. China

[5]Beijing Academy of Quantum Information Sciences, Beijing 100193, P. R. China

[†]These authors contributed equally to the work

*Correspondence to: xilin@pku.edu.cn (X.L.) and weihan@pku.edu.cn (W.H.)

**Abstract:**

Superconductivity has been one of the most fascinating quantum states of matter for over several decades. Among the superconducting materials, LaAlO$_3$/SrTiO$_3$ interface is of particularly interest since superconductivity exists between two insulating materials, which provides it with various unique applications compared with bulk superconductors and makes it a suitable platform to study the quantum Hall effect, charge density wave, superconductivity and magnetism in one device. Therefore, a lot of efforts have been made to search new superconducting oxide interface states with higher superconducting critical temperature ($T_C$). Recently, a superconducting state with $T_C \sim 2$ K has been found at the interface between a ferromagnetic insulator EuO and a band insulator (111)-KTaO$_3$. Here, we report the experimental investigation of the superconductor-metal quantum phase transition of the EuO/KTaO$_3$ interface. Around the transition, a divergence of the dynamical critical exponent is observed, which supports the quantum Griffiths singularity in the EuO/KTaO$_3$ interface. The quantum Griffiths singularity could be attributed to large rare superconducting regions and quenched disorders at the interface. Our results could pave the way for studying the exotic superconducting properties at the EuO/KTaO$_3$ interface.

## I. INTRODUCTION

Two-dimensional (2D) superconductivity at the LaAlO$_3$/SrTiO$_3$ interface has attracted a lot of attentions recently [1], which has exhibited interesting quantum phenomena, including the electrical-field-induced superconductor-insulator quantum phase transition [2] and coexistence of superconductivity with ferromagnetism [3-5]. Furthermore, LaAlO$_3$/SrTiO$_3$ interface might hold the promise towards future applications in the mesoscopic superconducting circuits [2]. Albeit its importance in fundamental physics and potential in applications, the extremely low $T_C$ (below 300 mK) is a critical challenge [2,6]. Very recently, unexpected superconductivity is observed at the EuO/(111)-KTaO$_3$ interface which shows a $T_C$ above 2 K [7]. The difference between 300 mK and 2 K is significant, because the former low temperature environment usually requires more expensive and rarer dilution refrigerators, while 2 K can be easily realized through evaporation of liquid helium or a close-cycle fridge simply based on electricity.

For 2D crystalline superconducting films/interfaces, the superconductor-metal/insulator phase transition is one of the most important properties [1]. Interestingly, in the case of quantum phase transitions involving a discrete symmetry breaking, i.e., when the system consists of large rare ordered regions, a quantum Griffiths singularity is theoretically expected to emerge [8,9]. Experimentally, the quantum Griffiths singularity has been observed in 2D superconducting Ga thin films [10], where the origin is theoretically discussed to be that the quenched disorder strongly influences the phase transition behavior and the resultant large rare superconducting regions are linked via long-range Josephson coupling. Subsequently, various low-dimensional superconducting systems have been shown to exhibit the quantum Griffiths singularity [11-16], albeit with detailed differences concerning the evolution of the dynamical critical exponent. Thus, testifying the quantum Griffiths singularity and the universality in a newly-discovered superconducting system is worth of more effort.

In this paper, we report the experimental investigation of superconductor-metal quantum phase transitions of the $EuO/(111)$-$KTaO_3$ interface. The superconductivity occurs at interface: the samples become as insulating as the pristine $KTaO_3$ substrates after etching the EuO. In the Hall bar geometry, the $T_C$ and $T_{BKT}$ are determined to be ~ 1.31 K and ~ 1.42 K, respectively, which are consistent with previous report utilizing the Van der Pauw geometry [7]. Around the superconductor-metal transition in the $EuO/KTaO_3$ interface, series of crossing points between neighboring magnetoresistance (MR) isotherms and the divergence of the dynamical critical exponent show up, which are features of quantum Griffiths singularity [10-16]. The quantum Griffiths singularity in the superconducting interface of the $EuO/KTaO_3$ heterostructures could arise from the quenched disorders that exist at the interface or the polycrystalline properties of the EuO layer.

## II. EXPERIMENTAL

The EuO/KTaO$_3$ heterostructures were prepared by growing EuO thin films on the (111)-KTaO$_3$ substrates via oxide molecular beam epitaxy (MBE-Komponenten GmbH; Octoplus 400). The (111)- KTaO$_3$ substrates were ordered from Hefei Kejing Material Technology Co., Ltd. Prior to the EuO growth, the KTaO$_3$ substrates were annealed at 500 ºC in vacuum (7 × 10$^{-10}$ mbar) for 30 min to clean the surface. Then the KTaO$_3$ substrates were kept at 500 ºC during the growth of EuO thin films (~ 10 nm), which included the following two steps. Firstly, a thin buffer layer of Eu (~ 1 nm) was grown onto the KTaO$_3$ substrate using a thermal effusion cell. Then, the oxygen gas with a pressure of 1×10$^{-9}$ mbar was induced into the chamber, and the EuO of ~ 9 nm was grown by reactive evaporation [17]. After the substrates were cooled down to ~ 50 ºC, a 5-nm MgO layer was deposited via *e*-beam evaporation before moving the samples out of the high-vacuum chamber. This thin MgO layer used to protect the EuO films from degradation during subsequent measurements.

The EuO/KTaO$_3$ Hall bar devices were fabricated using standard photolithography and wet-etching processes. Firstly, the EuO/KTaO$_3$ films were covered with photoresist by the spin coating process, and then were exposed to ultraviolet light through a photomask. Diluted hydrochloric acid was used to etch away the EuO to form the Hall bar geometry with a channel width of 100 μm and a length of 4500 μm. The last step was to remove the residual chemicals on devices via acetone, IPA and DI water subsequently.

For the electrical measurement from $T$ = 300 K to 2 K, the EuO/KTaO$_3$ heterostructures and Hall bar devices were measured in an Oxford Spectromag system with the d.c. technique ($I_{dc}$ ~ 100 μA). For the ultralow temperature measurement from 3.5 K to 71 mK, the resistance and magnetoresistance were measured in a dilution refrigerator (CF-CS81-600, Leiden Cryogenics BV) with the a.c. technique, using 10 ~ 100 nA excitation current at 17 Hz and 2 mT/s (Fig. 2(b)) or 1 mT/s (Fig. 3) sweeping rate. Home-made resistor-capacitor (RC) filters and silver-epoxy filters were used in the dilution refrigerator to filter external high-frequency radiation and lower the electron temperature of samples, respectively. The electron temperature in this fridge has been shown to be equal to the refrigerator temperature above 25 mK in previous fractional quantum Hall effect study [18].

## III. RESULTS

Figures 1(a) and (b) show the RHEED patterns of the (111)-KTaO$_3$ substrate and the polycrystalline EuO film (thickness: ~10 nm) viewed along [1-10] direction of KTaO$_3$. The polycrystalline nature for EuO grown on (111)-KTaO$_3$ substrates is due to the incompatible crystal structures and the large difference in in-plane lattice constant. The interface superconducting properties were characterized using the Van der Pauw (VdP) geometry (see Fig. S1 in supplementary materials) and Hall bar devices. For the VdP measurements with the current along the [11-2] and [1-10] directions, similar temperature dependences are observed, and $T_C$ is about 1.33 K determined from the zero-resistance temperature, and the onset superconducting temperature ($T_{C\_onset}$) is around 1.90 K.

The shcematics of the EuO/(111)-KTaO$_3$ Hall bar device are shown in Figure 1(c). After etching the EuO layer, the bare KTaO$_3$ part of the EuO/KTaO$_3$ heterostructures become insulating, which further confirms the interface superconductivity. A typical optical image of the EuO/KTaO$_3$ Hall bar device is shown in Fig. 1(d) inset. As the temperature decreases from 300 K to 1.5 K, the mobility increases from 5 cm$^2$/Vs to 98 cm$^2$/Vs, and the sheet carrier density decreases from ~2.4 × 10$^{14}$ cm$^{-2}$ to ~ 7.4 × 10$^{13}$ cm$^{-2}$ (supplementary Fig. S2). $T_C$, defined as the zero-resistance temperature, is determined to be ~1.31 K (Fig. 1(d)), which is almost the same as the EuO/KTaO$_3$ films characterized by Van der Pauw geometry.

Perpendicular magnetic field was applied to investigate the superconductor-metal transition of the EuO/KTaO$_3$ interface. As shown in Fig. 2(a), $T_C$ of the EuO/KTaO$_3$ Hall bar device is strongly suppressed by the magnetic field. Under the magnetic field of ~ 1.17 T, zero-resistance state cannot be reached down to $T$ = 71 mK, the lowest temperature of this measurement limited. When the magnetic field is further increased, the normal metallic state is induced. Based on the $B_{c\perp}$ vs. $T_C$ curve (supplementary materials and Fig. S3), the critical perpendicular magnetic field is estimated to be 1.51 T at the absolute zero temperature, which corresponds to a superconducting coherence length of ~ 14.8 nm. These results are consistent with previous report with those values of ~1.8 T and 13 nm at T = 0 K [7]. Figure 2(b) shows the longitudinal and Hall resistances as functions of the perpendicular magnetic field at $T$ = 91 mK. Clearly, a superconducting-to-normal-state transition happens at $B$ ~ 1.1 T. Above this critical magnetic field, both the longitudinal resistance and Hall signal exhibit normal-metallic behavior. Based on the linear Hall resistance signal as a function of the magnetic field, the sheet carrier density is obtained to be 7.3 × 10$^{13}$ cm$^{-2}$.

For 2D superconductors, Berezinskii-Kosterlitz-Thouless (BKT) transition characterizes the critical point where vortices and anti-vortices stabilize. When the temperature is slightly higher than $T_{BKT}$ but lower than the $T_{C\_onset}$, the vortices and anti-vortices are mobile, which results in finite resistances [19-21]. To determine the BKT temperature ($T_{BKT}$), the current-voltage curves are presented in Fig. 2(c) in log-log plot. As the temperature increases, the critical current decreases dramatically (inset of Fig. 2(d)). When the applied current is larger than the critical current, the current-voltage curves merge, showing a linear behavior similar to the normal metallic state at $T$ = 1.6 K. The power-law scaling of $V \propto I^3$ relationship is plotted (dashed purple line in Fig. 2(c)) to indicate the temperature around which the BKT transition happens [22]. To quantitatively determine $T_{BKT}$, the current-voltage curves are fitted by the following equation,

$$V \propto I^\alpha \qquad (1)$$

where α is the power-law coefficient. As the temperature decreases, α continuously increases from 1 at 1.60 K to 10 at 1.38 K (Fig. 2(d)), from which $T_{BKT}$ is determined to be 1.42 K.

Figure 3 shows the magnetoresistance isotherms measured on the EuO/KTaO$_3$ Hall bar device at various temperatures from 1.00 K to 0.080 K. Series of crossing points between neighboring MR isotherms are observed. As the temperature decreases, the magnetic field of the crossing points increases monotonically and a large enhancement of the critical magnetic field is observed at lower temperatures (inset of Fig. 3). The series of crossing points between MR isotherms are consistent with the quantum Griffiths singularity behaviors observed recently in 2D crystalline superconductors [10-13,15,16].

IV. DISCUSSION

To analyze this exotic phenomenon with series of crossing points, the finite-size scaling analysis is performed using the following formula [23-25],

$$R(B,T) = R_c f[(B - B_C)/T^{1/z\nu}], \qquad (2)$$

where $R_C$ is the critical resistance, $B_C$ is the critical magnetic field, $f[]$ is an arbitrary function with $f[0] = 1$, $z$ is the dynamic critical exponent, and $\nu$ is the coherence length exponent. For the purpose of effective analysis, the small crossing region with three adjacent $R(B)$ curves, is regarded approximately as one "critical" point. The Eq. (2) can be reformatted as $R(B_c,t)/R_c = f[(B-B_c)t]$, where $t = (T/T_0)^{-1/z\nu}$ and $T_0$ is the lowest temperature in each group [11]. For the temperature range of [0.8 K, 1.0 K], the critical values are $B_C = 1.483$ T, $R_C = 8263$ Ω, $T_0 = 0.8$ K. The finite-size scaling analysis of the MR isotherms is shown in fig. 4(a), and the data fall on to a bivalue curve. Similarly, figure 4(b) shows finite-size scaling analysis MR isotherms in the temperature range of [0.080 K, 0.125 K] with $B_C = 1.717$ T, $R_C = 8617$ Ω. The derived dynamical critical exponent ($z\nu$) is obtained to be 0.77 in high temperature range [0.80 K, 1.00 K] (inset of Fig. 4(a)) and 2.6 in the lowest temperature range [0.080 K, 0.125 K] (inset of Fig. 4(b)). The accuracy of critical exponent is about 10%. The systematical variation of $z\nu$ as a function of magnetic field for all the temperature ranges is summarized in Fig. 4(c). As the temperature decreases, $z\nu$ rapidly increases and exhibits a divergent behavior as a function of $B$. Quantitatively, this observation can be understood using the activated quantum scaling law [26,27],

$$z\nu \approx C|B - B_c^*|^{-\nu\psi}, \qquad (3)$$

where C is a constant, $B_c^*$ is the derived critical magnetic field, and $\nu$, $\psi$ are the 2D infinite randomness critical exponents with $\nu \sim 1.2$, and $\psi \sim 0.5$ [26,27]. The activated quantum scaling curve (dashed line in Fig. 4c) well describes the observed experimental results, and $B_c^*$ is obtained to be ~ 1.72 T.

The divergence of the dynamical critical exponent supports the quantum Griffiths singularity around the superconductor-metal transition, where singularity arises from the presence of large rare superconducting regions [28-30]. In this circumstance, the absence of long-range order does not necessarily blur the critical superconductor-metal quantum phase transition, but instead, the Griffiths singularity emerges due to the long-range Josephson coupling between separated superconducting islands, resulting global superconductivity [10]. It is also noted that recent theoretical studies [31,32] show that large rare superconducting regions are not sufficient for a true quantum Griffith singularity that is predicted by Fisher [8,9]. While experimentally, the superconductor to metal transitions in 2D Ga superconducting films exhibit the quantum Griffith behaviors [10], where the origin is theoretically attributed to the interplay of quenched disorder and thermal fluctuation. Our experimental results in EuO/KTaO$_3$ also exhibit the quantum Griffiths singularity feature around the superconductor-metal transition. Compared to previous reports, our observation is similar to quantum Griffith singularity reported in the 2D Ga superconducting films [10].

The observation of quantum Griffiths singularity in EuO/KTaO$_3$ indicates the possible quenched disorders that exist at the interface. Thus, the study of the correlation between the $T_C$ and the quantum Griffith singularity might be important to reveal the disorder effect on $T_C$ of the EuO/KTaO$_3$ interfaces. One of the possible sources of the quenched disorders could arise from the polycrystalline properties of the EuO layer, which is due to the large lattice mismatch between (111) EuO and (111)-faced KTaO$_3$. By improving the quality of the EuO thin films to remove the quenched disorders at the interface, the quantum Griffith singularity feature might disappear. Future works are needed to identify to what extent the extrinsic disorder can affect the quantum Griffiths singularity and the $T_C$ of the EuO/KTaO$_3$ interfaces.

## V. CONCLUSION

In summary, we have investigated the superconductor-metal transition properties of the interface superconductivity between a ferromagnetic insulator EuO and a band insulator $KTaO_3$. With a Hall bar geometry, the $T_C$ and $T_{BKT}$ are determined to be ~ 1.31 K and ~ 1.42 K, respectively, which are similar to the previous report of van der Pauw geometry measurement on the $EuO/KTaO_3$ interface [7]. Interestingly, the divergence of the dynamical critical exponent is observed as the temperature decreases, which supports the quantum Griffiths singularity in the $EuO/KTaO_3$ interface. The quantum Griffiths singularity could be attributed to large rare superconducting regions at the $EuO/KTaO_3$ interface. Our results bring motivation to further investigation of the exotic superconducting properties that could exist at the $EuO/KTaO_3$ interface, such as the coexistence of ferromagnetism and superconductivity, and unconventional spin-triplet superconductivity.

## ACKNOWLEDGMENTS

This work is supported by the National Basic Research Programs of China (Nos. 2019YFA0308401 and 2017YFA0303301), National Natural Science Foundation of China ((Nos. 11974025, 11674009, 11934016, and 11921004), Beijing Natural Science Foundation (No. 1192009 and JQ18002), and the Key Research Program of the Chinese Academy of Sciences (Grant No. XDB2800000).

**References:**

[1]  Y Saito, T Nojima, and Y Iwasa 2016 *Nat. Rev. Mater.* **2** 16094

[2] A D Caviglia, S Gariglio, N Reyren, D Jaccard, T Schneider, M Gabay, S Thiel, G Hammerl, J Mannhart, and J M Triscone 2008 *Nature (London)* **456** 624
[3] L Li, C Richter, J Mannhart, and R C Ashoori 2011 *Nat. Phys.* **7** 762
[4] J A Bert, B Kalisky, C Bell, M Kim, Y Hikita, H Y Hwang, and K A Moler 2011 *Nat. Phys.* **7** 767
[5] D A Dikin, M Mehta, C W Bark, C M Folkman, C B Eom, and V Chandrasekhar 2011 *Phys. Rev. Lett.* **107** 056802
[6] N Reyren, S Thiel, A D Caviglia, L F Kourkoutis, G Hammerl, C Richter, C W Schneider, T Kopp, A S Rüetschi, D Jaccard, M Gabay, D A Muller, J M Triscone, and J Mannhart 2007 *Science* **317** 1196
[7] X Y C. Liu, D. Jin, Y. Ma, H. Hsiao, Y. Lin, T. Sullivan, X. Zhou, J. Pearson, B. Fisher, J. Jiang, W.Han, J. Zuo, J. Wen, D. Fong, J. Sun, H. Zhou and A. Bhattacharya 2020 *arXiv* **2004** 07416
[8] D S Fisher 1992 *Phys. Rev. Lett.* **69** 534
[9] D S Fisher 1995 *Phys. Rev. B* **51** 6411
[10] Y Xing, H-M Zhang, H-L Fu, H Liu, Y Sun, J-P Peng, F Wang, X Lin, X-C Ma, Q-K Xue, J Wang, and X C Xie 2015 *Science* **350** 542
[11] S Shen, Y Xing, P Wang, H Liu, H Fu, Y Zhang, L He, X C Xie, X Lin, J Nie, and J Wang 2016 *Phys. Rev. B* **94** 144517
[12] Y Xing, K Zhao, P Shan, F Zheng, Y Zhang, H Fu, Y Liu, M Tian, C Xi, H Liu, J Feng, X Lin, S Ji, X Chen, Q-K Xue, and J Wang 2017 *Nano Lett.* **17** 6802
[13] Y Saito, T Nojima, and Y Iwasa 2018 *Nat. Commun.* **9** 778
[14] E Zhang, J Zhi, Y-C Zou, Z Ye, L Ai, J Shi, C Huang, S Liu, Z Lin, X Zheng, N Kang, H Xu, W Wang, L He, J Zou, J Liu, Z Mao, and F Xiu 2018 *Nat. Commun.* **9** 4656
[15] Y Liu, Z Wang, P Shan, Y Tang, C Liu, C Chen, Y Xing, Q Wang, H Liu, X Lin, X C Xie, and J Wang 2019 *Nat. Commun.* **10** 3633
[16] C Zhang, Y Fan, Q Chen, T Wang, X Liu, Q Li, Y Yin, and X Li 2019 *NPG Asia Mater.* **11** 76
[17] Y Yun, Y Ma, T Su, W Xing, Y Chen, Y Yao, R Cai, W Yuan, and W Han 2018 *Phys. Rev. Mater.* **2** 034201
[18] P Wang, K Huang, J Sun, J Hu, H Fu, and X Lin 2019 *Rev. Sci. Instrum.* **90** 023905
[19] V L Berezinskii 1971 *Sov. Phys. JETP* **32** 493
[20] V L Berezinskii 1972 *Sov. Phys. JETP* **34** 610
[21] Kosterli.Jm and D J Thouless 1972 *J. Phys. C* **5** 124
[22] K Epstein, A M Goldman, and A M Kadin 1981 *Phys. Rev. Lett.* **47** 534
[23] S L Sondhi, S M Girvin, J P Carini, and D Shahar 1997 *Rev. Mod. Phys.* **69** 315
[24] A M Goldman 2010 *Int. J. Mod. Phys. B* **24** 4081
[25] M P A Fisher 1990 *Phys. Rev. Lett.* **65** 923
[26] T Vojta, A Farquhar, and J Mast 2009 *Phys. Rev. E* **79** 011111
[27] I A Kovács and F Iglói 2010 *Phys. Rev. B* **82** 054437

[28]     N Markovic 2015 *Science* **350** 509
[29]     T Vojta and J A Hoyos 2014 *Phys. Rev. Lett.* **112** 075702
[30]     T Vojta 2006 *J. Phys. A* **39** R143
[31]     B Spivak, P Oreto, and S A Kivelson 2008 *Phys. Rev. B* **77** 214523
[32]     A Kapitulnik, S A Kivelson, and B Spivak 2019 *Rev. Mod. Phys.* **91** 011002
[33]     C Richter, H Boschker, W Dietsche, E Fillis-Tsirakis, R Jany, F Loder, L F Kourkoutis, D A Muller, J R Kirtley, C W Schneider, and J Mannhart 2013 *Nature (London)* **502** 528
[34]     B Kalisky, E M Spanton, H Noad, J R Kirtley, K C Nowack, C Bell, H K Sato, M Hosoda, Y Xie, Y Hikita, C Woltmann, G Pfanzelt, R Jany, C Richter, H Y Hwang, J Mannhart, and K A Moler 2013 *Nat. Mater.* **12** 1091

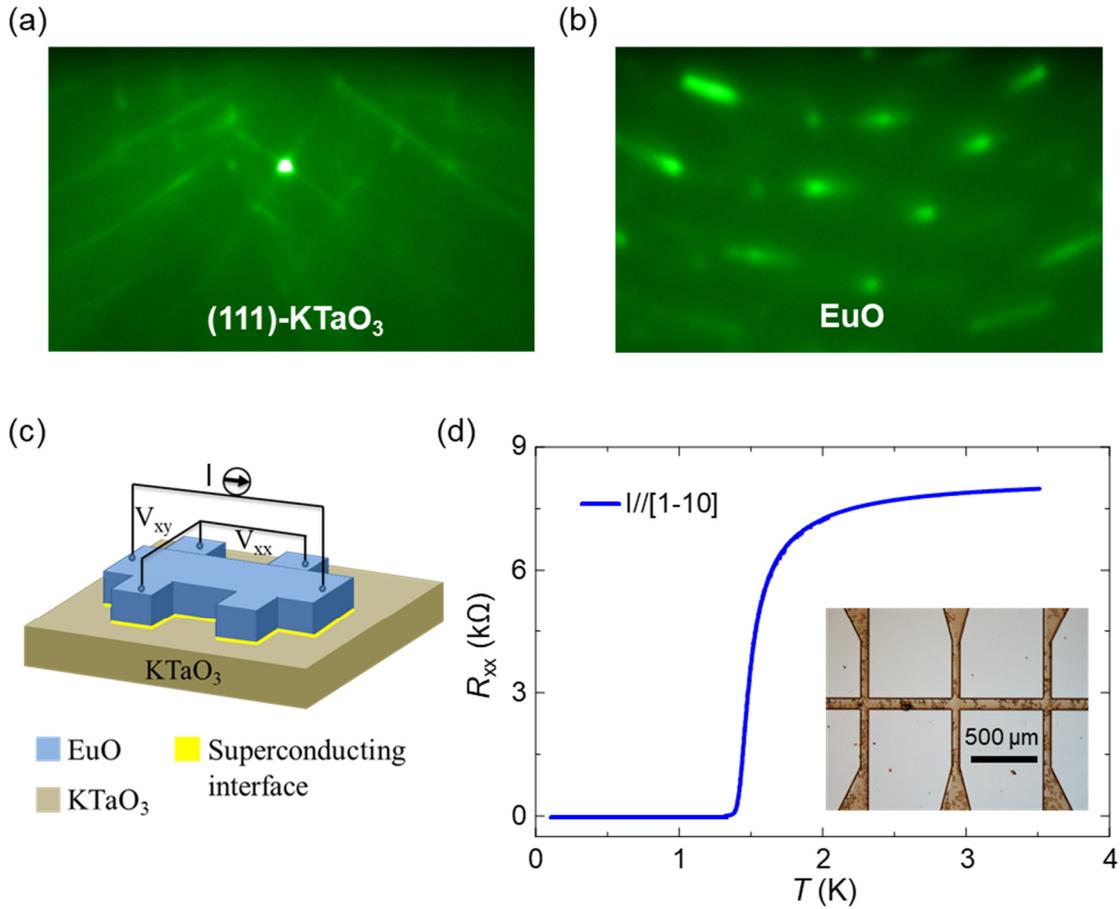

**Figure 1. Growth and electrical measurement of the superconducting EuO/(111)-KTaO$_3$ interface.** (a-b) RHEED patterns of a typical (111)-oriented KTaO$_3$ substrate and the polycrystalline EuO thin film (~10 nm) viewed from the KTaO$_3$ crystal's [1-10] direction. (c) Schematic of the EuO/KTaO$_3$ Hall bar device and measurement geometry. (d) The longitudinal resistance ($R_{xx}$) as a function of temperature on the EuO/KTaO$_3$ Hall bar device. Inset: The optical image of the Hall bar device.

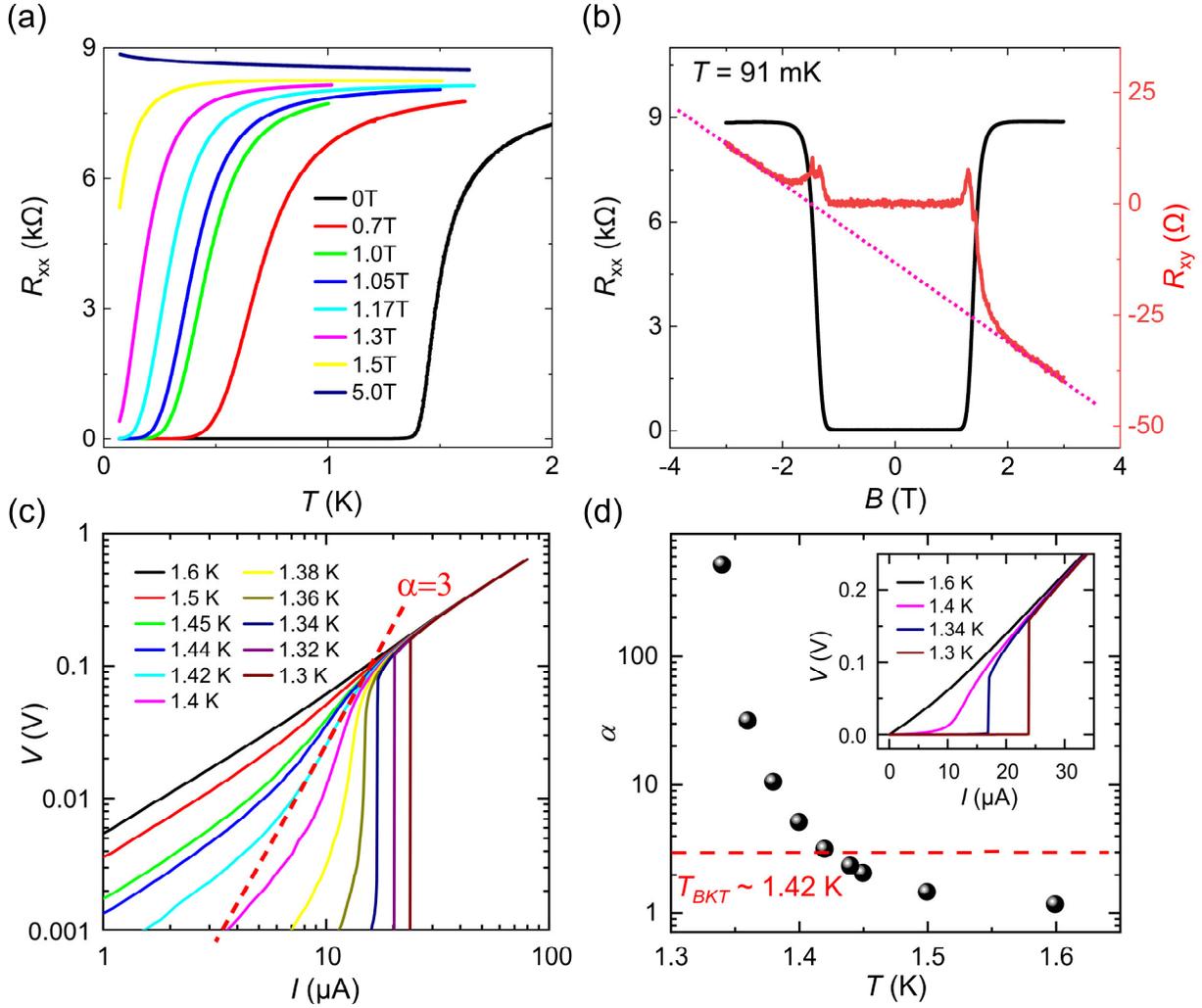

**Figure 2. The BKT transition of the superconducting EuO/(111)-KTaO$_3$ interface.** (a) The temperature dependence of the longitudinal resistance of the EuO/KTaO$_3$ Hall bar device under various magnetic fields. The magnetic field is applied perpendicular to the interface. (b) The magnetic field dependence of the longitudinal and Hall resistances of the EuO/KTaO$_3$ Hall bar device at $T$ = 91 mK. The purple dashed line represents the best linear fit for the Hall resistance at large magnetic fields. (c) The log-log curves of current-voltage dependence between $T$ = 1.3 and 1.6 K. The red dashed line represents the relationship of $V \sim I^3$. (d) The BKT transition temperature ($T_{BKT}$) determined from the power coefficient ($\alpha$) vs. $T$ plot. Inset: The current-voltage curves at $T$ = 1.3, 1.34, 1.4, and 1.6 K, respectively.

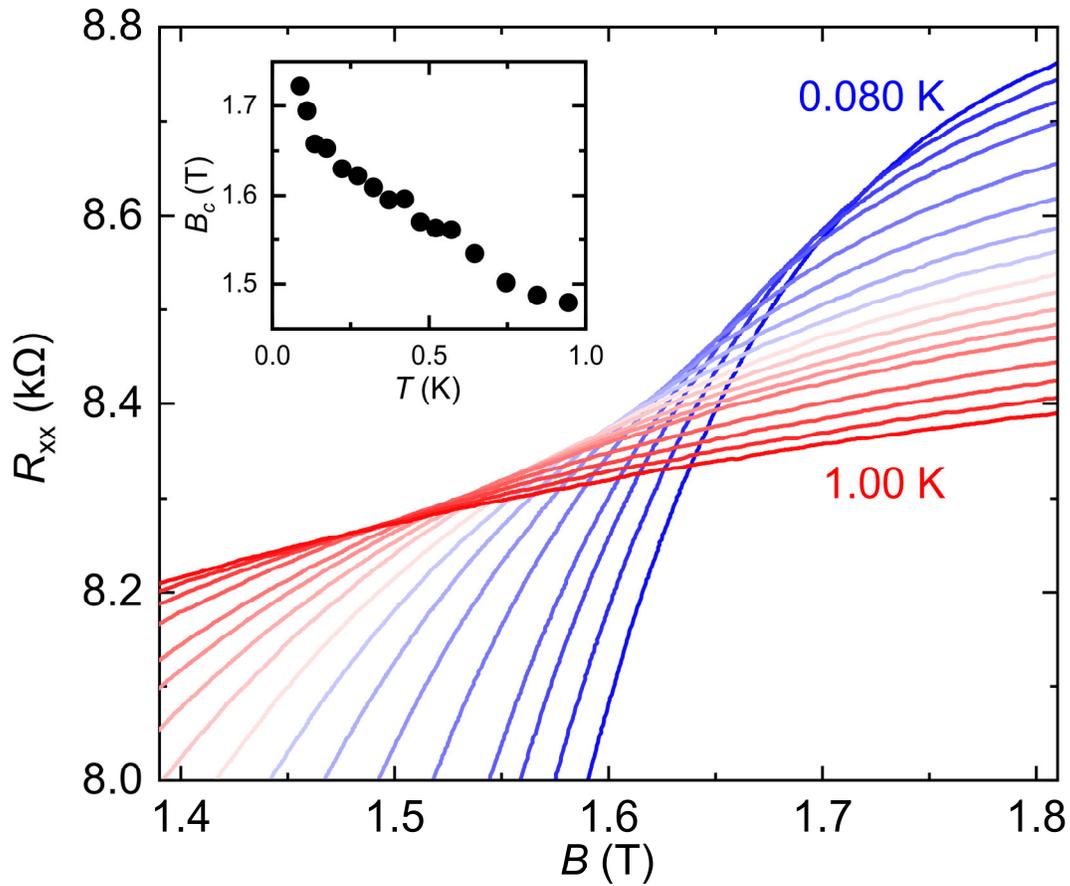

**Figure 3. Magnetoresistance isotherms of the superconducting EuO/KTaO$_3$ interface.** The magnetic field dependence of the longitudinal resistance of the EuO/KTaO$_3$ Hall bar device at various temperatures. Inset: The temperature dependence of the crossing magnetic field ($B_c$) that could be extracted from $R_{xx}$ vs. $B$ curves at every two adjacent temperatures.

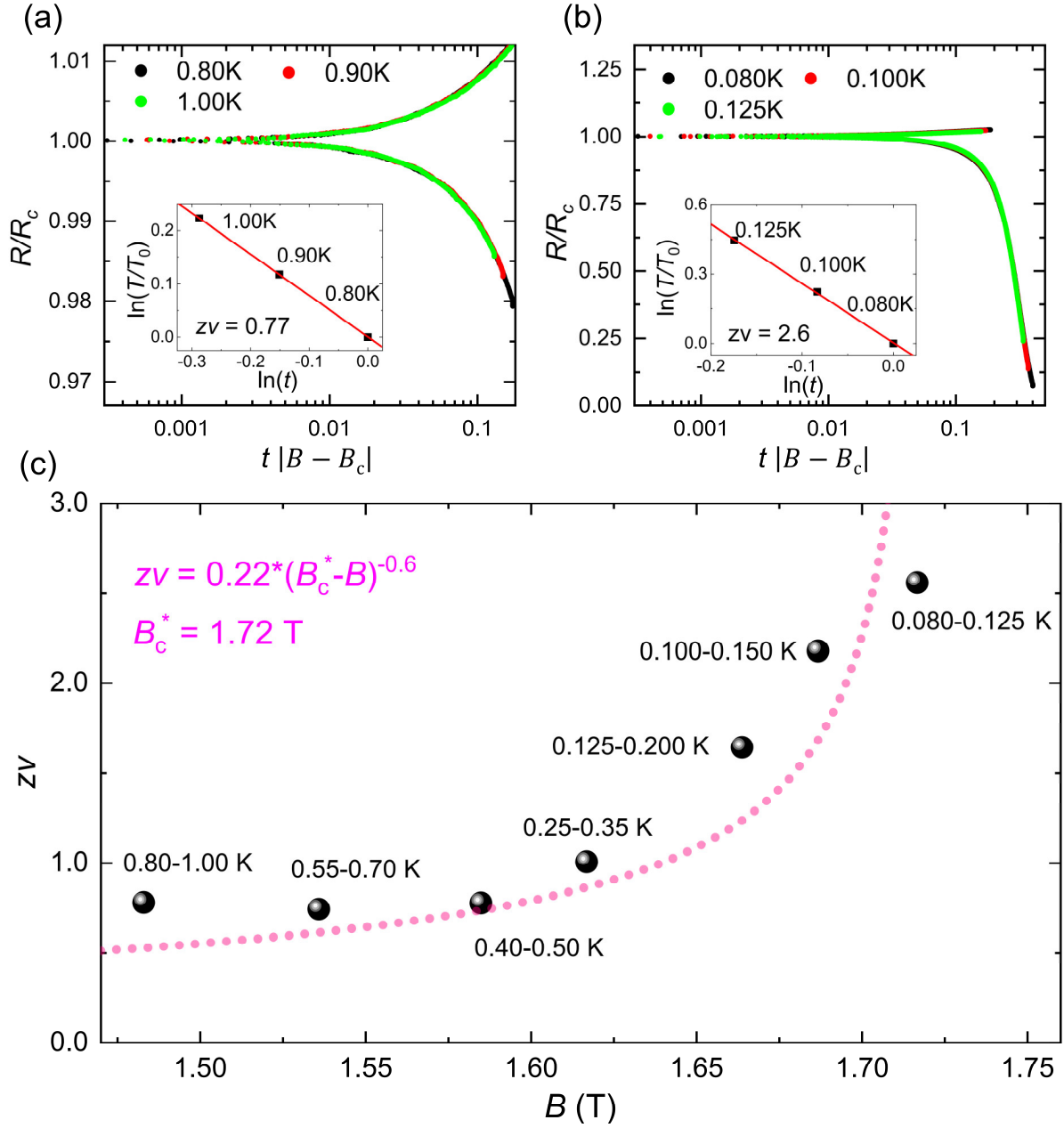

**Figure 4. Experimental evidence of quantum Griffiths singularity at the EuO/KTaO$_3$ interface.** (a-b) The scaling analysis of the interface superconductivity for two typical temperature regimes (0.80 K - 1.00 K and 0.080 K - 0.125 K). (c) The magnetic field dependence of dynamical critical exponent ($zv$) in various temperature regimes. The purple dashed line represents the best fit based on the equation $zv = 0.22 \times (B_c^* - B)^{-0.6}$.

# Supplementary Materials for

# Superconductor-metal quantum transition at the EuO/KTaO$_3$ interface

Yang Ma[1†], Jiasen Niu[1†], Wenyu Xing[1], Yunyan Yao[1], Ranran Cai[1], Jirong Sun[2,3], X. C. Xie[1,4,5], Xi Lin[1,4,5]*, Wei Han[1]*

[1]International Center for Quantum Materials, School of Physics, Peking University, Beijing 100871, P. R. China

[2]Beijing National Laboratory for Condensed Matter Physics & Institute of Physics, Chinese Academy of Sciences, Beijing 100190, P. R. China

[3]School of Physical Sciences, University of Chinese Academy of Sciences, Beijing 100049, P. R. China

[4]CAS Center for Excellence in Topological Quantum Computation, University of Chinese Academy of Sciences, Beijing 100190, P. R. China

[5]Beijing Academy of Quantum Information Sciences, Beijing 100193, P. R. China

†These authors contributed equally to the work

*Correspondence to: xilin@pku.edu.cn (X.L.) and weihan@pku.edu.cn (W.H.)

**Supplementary Note 1. Estimation of the coherence length of the superconducting EuO/KTaO$_3$ interface.**

The coherence length ($\xi_{GL}$) at absolute zero tempeature is calculated using the following equation,

$$\xi_{GL} = \sqrt{\frac{\Phi_0}{2\pi B_{C\perp(T\to 0)}}}, \qquad (S1)$$

where $\Phi_0$ is the quantum flux, $B_{C\perp(T\to 0)}$ is the critical perpendicular magnetic field at absolute zero tempeature. $B_{C\perp}$ at each tempeature can be obtained form the half value of the normal resitance during the superconductor-metal transition. Based on the $B_{C\perp}$ vs. $T$ curve (Fig. S2), the critical magnetic field at zero temperature is determined to be ~ 1.51 T, and the coherence length is calculated to be ~ 14.8 nm.

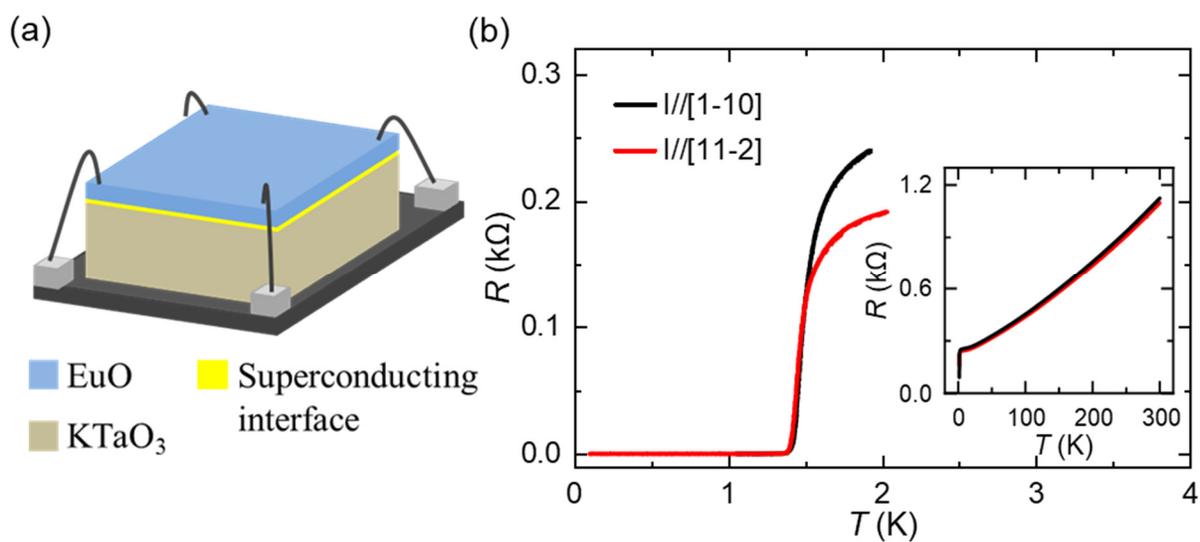

**Figure S1. The transport properties of the EuO/KTaO$_3$ interface in the VdP geometry.**
(a) The schematic illustration of the van der Pauw measurement geometry of the EuO/KTaO$_3$ interface. (b) The electrical measurement of the interface superconductivity with the current along the KTaO$_3$ substrate's [1-10] and [11-2] directions. Inset: The measured resistance as a function of temperature from 300 K to 1.5 K.

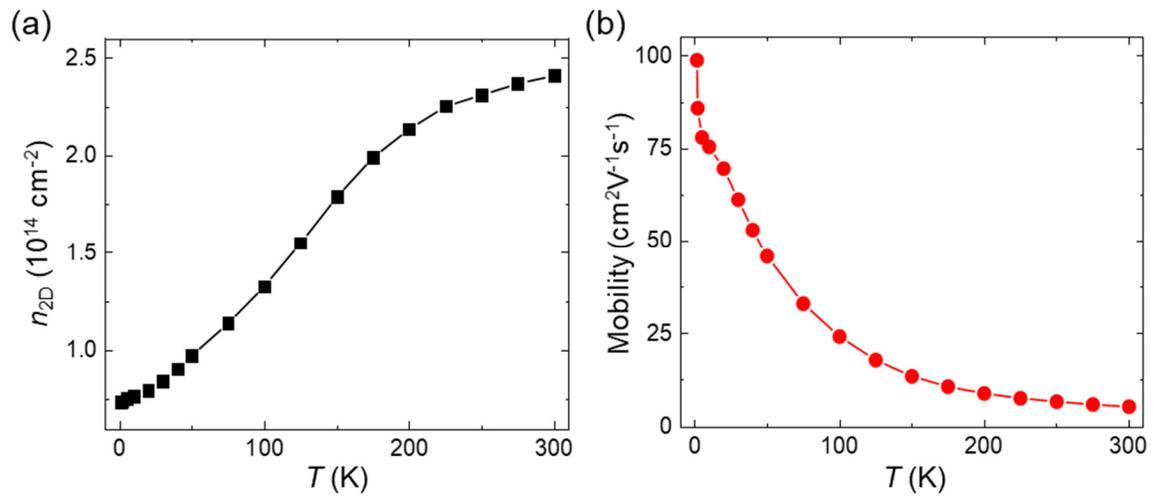

**Figure S2. The electron transport properties of the EuO/KTaO$_3$ interface above $T_C$.**
(a-b) The sheet carrier density and electron mobility as a function of temperature.

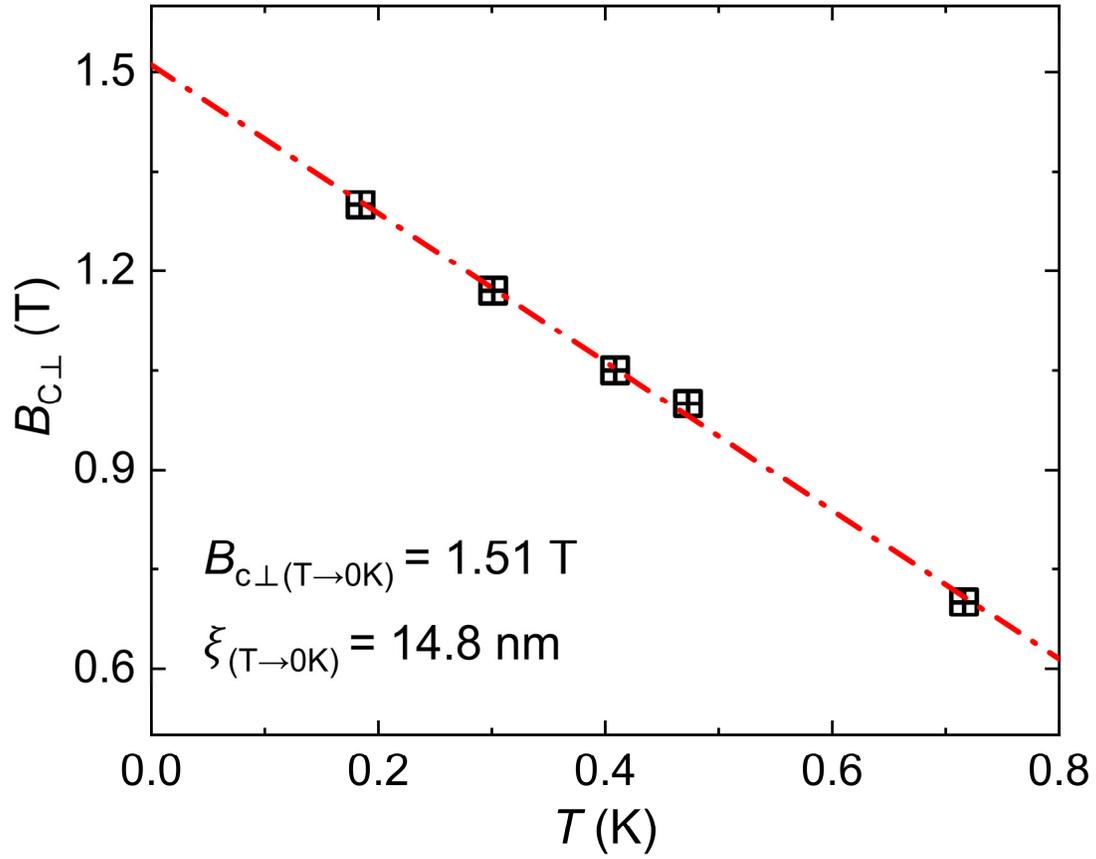

**Figure S3. Temperature dependence of the critical perpendicular magnetic field of the EuO/KTaO₃ interface.** The red line represents the best linear fit.